\documentclass[twocolumn,epjc3]{svjour3}
\RequirePackage[T1]{fontenc}
\RequirePackage{amssymb}
\RequirePackage{bm}
\RequirePackage{flushend}

\RequirePackage{float}
\smartqed  
\RequirePackage{graphicx}
\RequirePackage{mathptmx}      
\RequirePackage{latexsym}
\RequirePackage[numbers,sort&compress]{natbib}

\AtBeginDocument{}
\journalname{Eur. Phys. J. C}

\begin{document}

\title{Probing Shadowed Nuclear Sea with Massive Gauge Bosons in the Future Heavy-Ion Collisions
}


\author{Peng Ru\thanksref{e1,addr1,addr2}
        \and
        Ben-Wei Zhang\thanksref{e2,addr2}
        \and
        Enke Wang\thanksref{addr2}
        \and
        Wei-Ning Zhang\thanksref{addr1}}

\thankstext{e1}{e-mail: pengru@mail.dlut.edu.cn}
\thankstext{e2}{e-mail: bwzhang@mail.ccnu.edu.cn}


\institute{School of Physics $\&$ Optoelectronic Technology, Dalian
University of Technology, Dalian, 116024, China \label{addr1}
           \and
           Key Laboratory of Quark \& Lepton Physics (MOE) and Institute of Particle Physics,
 Central China Normal University, Wuhan, 430079, China \label{addr2}
}

\date{Received: date / Accepted: date}

\maketitle

\begin{abstract}
The production of the massive bosons $Z^0$ and $W^{\pm}$ could provide an excellent tool to study cold nuclear matter effects and the modifications of nuclear parton distribution functions (nPDFs) relative to parton distribution functions (PDFs) of a free proton in high energy nuclear reactions at the LHC as well as in heavy-ion collisions (HIC) with much higher center-of mass energies available in the future colliders.
In this paper we calculate the rapidity and transverse momentum distributions of the vector boson and their nuclear modification factors in p+Pb collisions at $\sqrt{s_{NN}}=63$~TeV  and in Pb+Pb collisions at $\sqrt{s_{NN}}=39$~TeV in the framework of perturbative QCD by utilizing three parametrization sets of nPDFs: EPS09, DSSZ and nCTEQ.
It is found that in heavy-ion collisions at such high colliding energies, both the rapidity distribution and the transverse momentum spectrum of vector bosons are considerably suppressed in wide kinematic regions with respect to p+p reactions due to large nuclear shadowing effect.
We demonstrate that in the massive vector boson productions processes with sea quarks in the initial-state may give more contributions than those with valence quarks in the initial-state, therefore in future heavy-ion collisions the isospin effect is less pronounced and the charge asymmetry of W boson will be reduced significantly as compared to that at the LHC.  Large difference between results with nCTEQ and results with EPS09 and DSSZ is observed in nuclear modifications of both rapidity and $p_T$ distributions of $Z^0$ and $W$ in the future HIC.


~
~

\end{abstract}

\section{Introduction}
\label{introduction}
The production of massive vector bosons $Z^0/W^\pm$ in high-energy nuclear collisions has long been regarded as an excellent tool to probe the initial-state cold nuclear matter~(CNM) effects
in the ultra-relativistic heavy-ion collisions (HIC) at the LHC because $Z^0/W^\pm$ do not participate in strong interactions and the mean-free-path of massive vector bosons (or more precisely, the leptons decayed from $Z^0/W^\pm$ ) is rather long ~\cite{Aad:2010aa,Chatrchyan:2011ua,Chatrchyan:2012nt,Aad:2012ew,atlasz14,CMS:2014kla,Aaij:2014pvu,Chatrchyan:2014csa,Neufeld:2010dz,
Neufeld:2010fj,Ru:2014yma}. In proposed heavy-ion programs at the Future Circular Collider (FCC) at CERN~\cite{FCC}, and the Circular Electron Positron Collider
 with a subsequent Super proton-proton Collider (CEPC-SPPC) in China~\cite{CEPC}, the vector gauge boson productions may still play a very important role in investigating the high-density QCD in the initial-state and making precise constraints on CNM effects and nuclear parton distribution functions (nPDFs) at very small $x$.  We note that the massive vector bosons are produced very early at time $\sim 1/m_{Z,W}\sim 10^{-3}$~fm/c with a decay time smaller than $0.1$~fm/c, and a mean-free-path $\sim10$~fm in the QGP at temperature $1$~GeV~\cite{ConesadelValle:2009vp}. Even though the hot/dense QCD matter to be created
in nucleus-nucleus collisions at the FCC and the CEPC-SPPC may have a much higher initial temperature and longer lifetimes~\cite{Armesto:2014iaa},
the vector boson with the final state colorless di-lepton would still be nearly blind to the medium
evolution~\cite{Kartvelishvili:1995fr,ConesadelValle:2007sw,ConesadelValle:2009vp,Xing:2011fb}.

Recent studies have shown that the heavy gauge boson production in nuclear reactions at the LHC energies could shed light on several
CNM effects, especially the shadowing and anti-shadowing effects~(for $Z/W$) and
isospin effect~(for $W^\pm$)~\cite{Vogt:2000hp,Paukkunen:2010qg,Guzey:2012jp,Kang:2012am,Albacete:2013ei,Ru:2014yma}.
In the conceptual designs of FCC~\cite{Armesto:2014iaa} and CEPC-SPPC~\cite{CEPC-preCDR}
the center-of-mass energy for proton-proton collisions  could reach up to $100$~TeV, which may give the energy of about $\sqrt{s_{NN}}=63$~TeV for proton-lead collisions and about $\sqrt{s_{NN}}=39$~TeV for lead-lead collisions.
With much higher colliding energies available in future HIC, the vector boson would be produced by initial partons with the much smaller momentum fraction $x$, in which region the shadowing effect will be more pronounced and sea quarks and gluons may play a more important role. It is of great interest to see how the massive gauge boson productions help us understand the CNM effects and impose stringent constraints on nPDFs in future heavy-ion collisions. In this paper we study the nuclear modifications of the vector boson production in the future Pb+Pb
collisions at $\sqrt{s_{NN}}=39$~TeV and p+Pb collisions at $\sqrt{s_{NN}}=63$~TeV.
The numerical calculations are performed by using the perturbative quantum chromo-dynamics~(pQCD) program DYNNLO~\cite{DYNNLO} incorporating
the parameterized nuclear parton distribution functions~(nPDFs) sets EPS09~\cite{Eskola:2009uj}, DSSZ~\cite{deFlorian:2011fp} and nCTEQ~\cite{Schienbein:2009kk,Kovarik:2010uv}.

This paper is organized as follows.
In Section~\ref{section:Zrapidity}, we briefly introduce the framework of our calculation and then compute
the nuclear modification ratios for $Z^0$ boson rapidity distributions and charged lepton pseudo-rapidity dependence of $W^\pm$ production in Pb+Pb and p+Pb collisions in the future HIC.
In Section~\ref{section:ZWtransverse}, we discuss the nuclear modification of the transverse momentum distributions
of $Z^0$ and $W^\pm$ bosons in the future HIC.
We present our summary in Section~\ref{section:summary&conclusion}.

\section{Nuclear modification ratio for vector boson rapidity distributions}
\label{section:Zrapidity}

\begin{figure}[t]
\begin{center}
\includegraphics[scale=0.70]{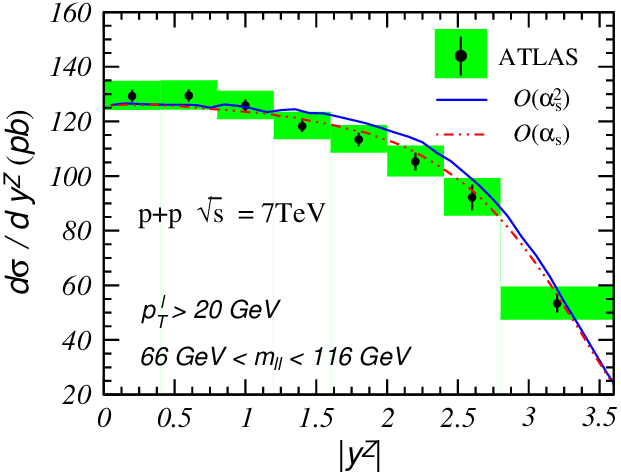}
\end{center}
\caption{(Color online) The rapidity distribution of $Z^0$ boson in p+p collisions at $\sqrt{s}=7$~TeV.
The ATLAS combined data are taken from Ref.~\cite{Aad:2011dm}. }
\label{fig:Z-y@pp}
\end{figure}

Before discussing the nuclear modification of vector boson production,
we briefly introduce the theoretic framework of this study,
which is the same as in reference~\cite{Ru:2014yma}.
In hadronic collisions, the differential cross section of vector boson production $A+B\rightarrow V+X\rightarrow ll+X$ through the Drell-Yan mechanism~\cite{DY},
could be calculated within the QCD-improved parton model~\cite{Fie95, QCDcp}. In this approach, the cross section could be factorized into the convolution of the parton distribution functions~(PDFs) $f(x,\mu)$ and partonic cross section $\hat{\sigma}_{ab\to V+X\to ll+X}$ which could be computed with the perturbative expansion,
\begin{eqnarray}
\label{DY}
\sigma_{DY}&=&\sum_{a,b}\int dx_ad x_b f_{a/A}(x_a,\mu)f_{{b/B}}(x_b,\mu) \nonumber \\
&\times&[\hat{\sigma}_0+\frac{\alpha_s}{2\pi}\hat{\sigma}_1+\left(\frac{\alpha_s}{2\pi}\right)^2\hat{\sigma}_2+\cdots]_{ab\to V+X\to ll+X}, \nonumber \\
\end{eqnarray}
where $x_{a(b)}$ are the momentum fractions of the parton $a(b)$ from the hadron $A(B)$.
As the baseline of the study, the hadron-hadron cross sections are numerically calculated with the program DYNNLO~\cite{DYNNLO}, which
provides the $\mathcal {O}(\alpha_s)$ and $\mathcal {O}(\alpha_s^2)$ pQCD corrections to the
leading-order~(LO) cross section $[\hat{\sigma}_0]_{q\bar{q}\to V\to ll}$.
The $\mathcal {O}(\alpha_s)$ calculation includes real corrections and one-loop virtual corrections, and that at $\mathcal {O}(\alpha_s^2)$ includes
double-real corrections, real-virtual corrections, as well as two-loop virtual corrections~\cite{DYNNLO}.
In addition, the parton distribution functions parametrization sets MSTW20-08~\cite{Martin:2009iq} are used,
and the renormalization and factorization scales are set at the boson mass $m_V$.
\begin{figure}[t]
\begin{center}
\includegraphics[scale=0.70]{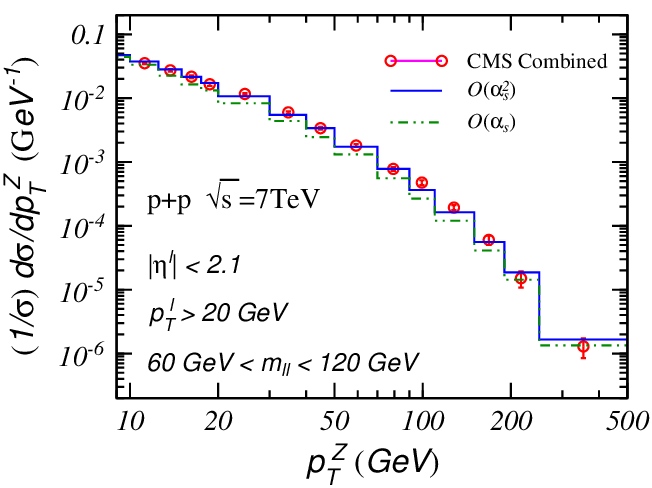}
\end{center}
\caption{(Color online) The normalized $Z^0$ boson transverse momentum spectrum in p+p collisions at $\sqrt{s}=7$~TeV.
The CMS combined data are taken from Ref.~\cite{Chatrchyan:2011wt}. }
\label{fig:Z-pt@pp}
\end{figure}

\begin{figure*}[t]
\includegraphics[scale=0.7]{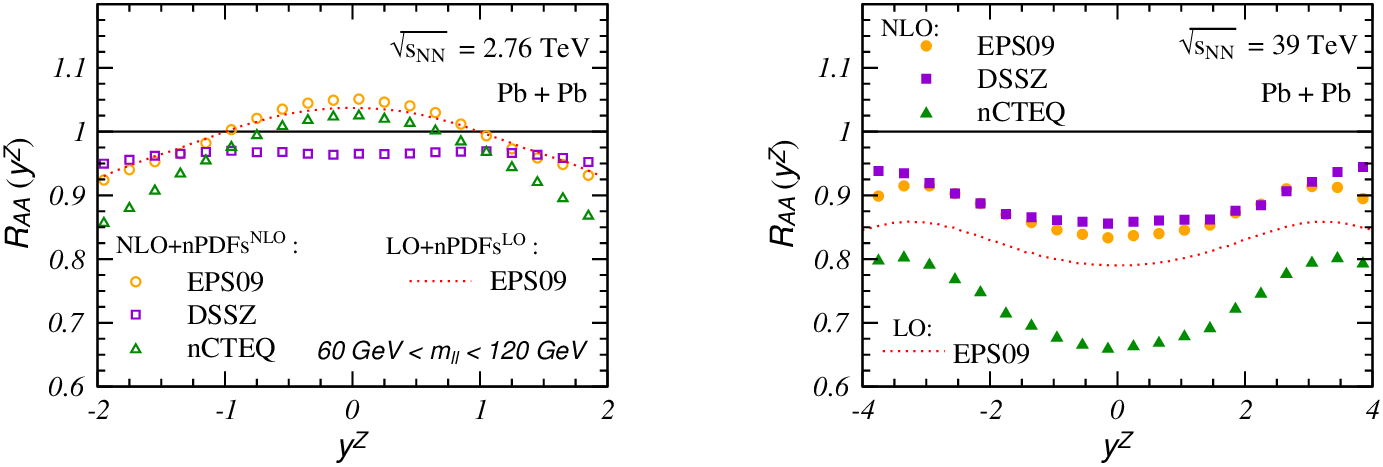}
\caption{(Color online) The nuclear modification ratio $R_{AA}$ of the $Z^0$ boson rapidity dependence for
$\sqrt{s_{NN}}=2.76$~TeV~(LHC, left panel) and $\sqrt{s_{NN}}=39$~TeV~(future HIC, right panel) at the NLO.
The LO results with EPS09 LO nPDFs are also plotted for comparison.}
\label{fig:Z-raay@PbPb}
\end{figure*}

We plot in Fig.\ref{fig:Z-y@pp} the $Z^0$ rapidity distribution,
and in Fig.\ref{fig:Z-pt@pp} its transverse momentum spectrum,
for proton-proton collisions at $\sqrt{s}=7$~TeV.
For rapidity dependence, both $\mathcal {O}(\alpha_s)$ and $\mathcal {O}(\alpha_s^2)$ pQCD calculations agree well with
the ATLAS data~\cite{Aad:2011dm}, which is measured in the invariant mass interval of the lepton pair as $66$~GeV$< m_{ll}<116$~GeV and also in the lepton transverse momentum region as $p_T^l>20$~GeV.
The calculation at $\mathcal {O}(\alpha_s^2)$ also give good description of the CMS transverse momentum distribution~\cite{Chatrchyan:2011wt} in the region $p_T^Z>10$~GeV. The CMS measurement is made in the final state phase volume defined by the invariant mass,
transverse momentum, and pseudo-rapidity of the dilepton as: $60$~GeV$< m_{ll}<120$~GeV, $p_T^l>20$~GeV, and $|\eta^l|<2.1$.
The DYNNLO program has also been shown to give good agreement with the CMS data on $Z^0$ rapidity distribution,
and with the transverse momentum spectra of $Z^0$ and $W$ measured by ATLAS~\cite{Ru:2014yma}.

To include several important CNM effects in Pb+Pb and p+Pb collisions,
we phenomenologically utilize the parametrization sets of nuclear parton distribution functions~(nPDFs)~\cite{Eskola:2009uj,deFlorian:2011fp,Schienbein:2009kk,Kovarik:2010uv,Dai:2013xca,He:2011sg,Dai:2012am}.
In this work, we chose MSTW2008 proton PDFs and multiply them by the flavor and scale dependent factors $R_f(x,\mu)$
taken from EPS09, DSSZ, and nCTEQ to obtain the parton distribution $f^{p,A}(x,\mu)$ of a bound proton in nucleus.
With the isospin symmetry being assumed~\cite{Eskola:2009uj,deFlorian:2011fp,Schienbein:2009kk,Kovarik:2010uv}, the nuclear parton distribution of a bound neutron $f^{n,A}(x,\mu)$ could be obtained. In this paper
only minimum-bias~(MB) nuclear collisions are considered.

Then we focus on the nuclear modification on vector boson production in the future HIC.
First we study that on $Z^0$ rapidity distribution in the Pb+Pb collisions at $\sqrt{s_{NN}}=39$~TeV.
The $Z^0$ signal is defined in the invariant mass window of the final state lepton pair as $60$~GeV$<m_{ll}<120$~GeV,
which is the same as the CMS experiments~\cite{Chatrchyan:2011ua,Chatrchyan:2014csa}.
The $Z-\gamma^*$ interference is included in the calculation.
We calculate at the order $\mathcal {O}(\alpha_s)$, the nuclear modification ratio $R_{AA}(y^Z)$ defined as
\begin{eqnarray}
R_{AA}(y^Z)=\frac{d\sigma^{AA}/dy^Z}{\langle N_{coll}\rangle d\sigma^{pp}/dy^Z}. \, \,
\end{eqnarray}
with $\langle N_{coll}\rangle$ the number of binary nucleon-nucleon collisions~\cite{d'Enterria:2003qs}.

In Fig.~\ref{fig:Z-raay@PbPb} both the results at the LHC and future energies are plotted for comparison.
Significant differences between $R_{AA}(y^Z)$ at two energies can be seen.
The LHC results~(left panel) with three nPDFs sets show weak nuclear modifications~($\lesssim10\%$) in the studied rapidity region.
Enhancements with EPS09 and nCTEQ at mid-rapidity can be observed, where suppression is given by DSSZ.
 However, at future HIC~(right panel), strong suppressions can be found in the whole studied rapidity regime~($|y|<4.0$).
Especially, nCTEQ nPDFs may suppress more than $30\%$ of $Z^0$ yield at the central rapidity, relative to proton-proton collisions.
EPS09 and DSSZ give similar nuclear variations for the future HIC.
The LO results with EPS09 are also plotted in Fig.~\ref{fig:Z-raay@PbPb}.
It is observed that the higher order corrections make negligible alteration on the $R_{AA}$ at the LHC,
but give a larger suppression on $R_{AA}$ at future HIC.

\begin{figure}[t]
\includegraphics[scale=0.7]{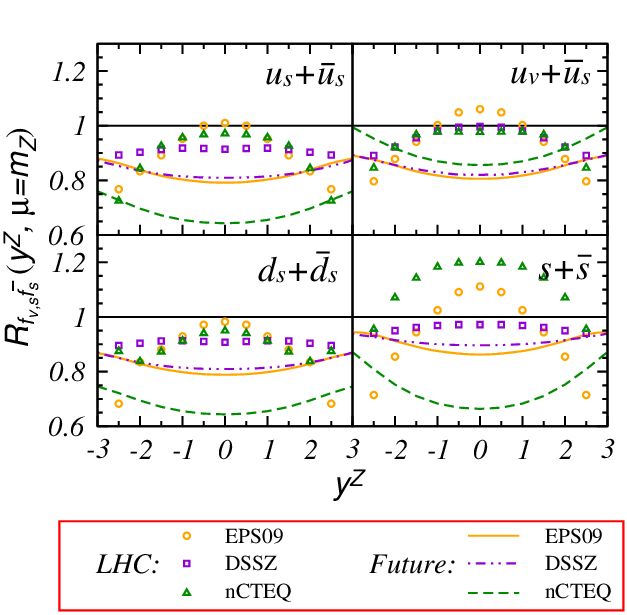}
\caption{(Color online) The subprocess dependent factor $R_{f_{v,s}\bar{f}_s}(y^Z,\mu)$ for Pb+Pb collisions
 at both $\sqrt{s_{NN}}=2.76$~TeV and $39$~TeV with factorization scale set at $m_Z$.}
\label{fig:Z-rffbar@Pb+Pb}
\end{figure}

To give a simple picture of the difference between the two collisions energies,
we resort to a lowest-order analysis~\cite{Ru:2014yma}.
In the LO pure electro-weak processes, the rapidity of $Z^0$ is related to the initial quark and anti-quark
momentum fractions with the kinematic equation
\begin{eqnarray}
\label{xy}
x_{1,2}=\frac{m_Z}{\sqrt{s_{NN}}}e^{\pm y^Z},
\end{eqnarray}
where the narrow width approximation~($m_{ll}\approx m_Z$) is used.
Equation~(\ref{xy}) shows that the momentum fraction of the initial parton is inversely proportional
to the colliding energy $\sqrt{s_{NN}}$.
For example, one has $x\sim0.033$ for mid-rapidity $Z^0$ production at $2.76$~TeV,
which falls into the anti-shadowing region of EPS09, whereas at $39$~TeV $x\sim0.0023$ is given by Eq.~(\ref{xy}),
that is small enough to enter the shadowing region of EPS09. The suppression in the $39$~TeV Pb+Pb collisions
is due to the shadowed smaller-$x$ initial partons.
To visually display the difference between the two energies, we plot in Fig.~\ref{fig:Z-rffbar@Pb+Pb}
the factor $R_{f_{v,s}\bar{f}_s}(y^Z,\mu)$ written as
\begin{eqnarray}
\label{rffbary}
R_{f_{v,s}\bar{f}_s}(y^Z,\mu)=&\frac{1}{2}&[R_{f_{v,s}}(x_1,\mu)R_{\bar{f}_s}(x_2,\mu) \nonumber \\
&+&R_{f_{v,s}}(x_2,\mu)R_{\bar{f}_s}(x_1,\mu)] \,\, ,
\end{eqnarray}
where $R_f(x,\mu)$ is the flavor dependent nuclear variation factor given by EPS09, DSSZ, or nCTEQ,
and $x_{1,2}$ is related to $y^Z$ by Eq.~(\ref{xy}).
The subscripts $v$ and $s$ stand for the valence and sea quarks~(no gluon at leading order), respectively.
This $R_{f_{v,s}\bar{f}_s}(y^Z,\mu)$ factor could reflect the nuclear modification on the LO partonic subprocess
initiated with $f_{v,s}$ and $\bar{f}_s$, thus could shed light on the $R_{AA}(y^Z)$~\cite{Ru:2014yma}.
For the four processes plotted in Fig.~\ref{fig:Z-rffbar@Pb+Pb}, similar differences between the two collision energies could be observed.
Compared to those at the LHC, the PDFs in future Pb+Pb collisions are more depleted in the kinematic region of the mid-rapidity $Z^0$ production, which results in a larger suppression of $R_{AA}(y^Z)$ in the future HIC than that at the LHC as observed in Fig.~\ref{fig:Z-raay@PbPb}.
We also notice that, although at the future HIC the $u_v+\bar{u}_s$ channel with nCETQ gives less suppression than those with EPS09 and DSSZ, their contribution combined with relatively large suppressions by other channels with nCTEQ, may give a more suppressed $R_{AA}(y^Z)$ than those with EPS09 and DSSZ. The underlying reason is that the  $u_v+\bar{u}_s$ channel gives relatively small contribution to $Z^0$ yield in the future HIC, and we will discuss this issue in more detail in the following.

\begin{figure}[t]
\includegraphics[scale=0.7]{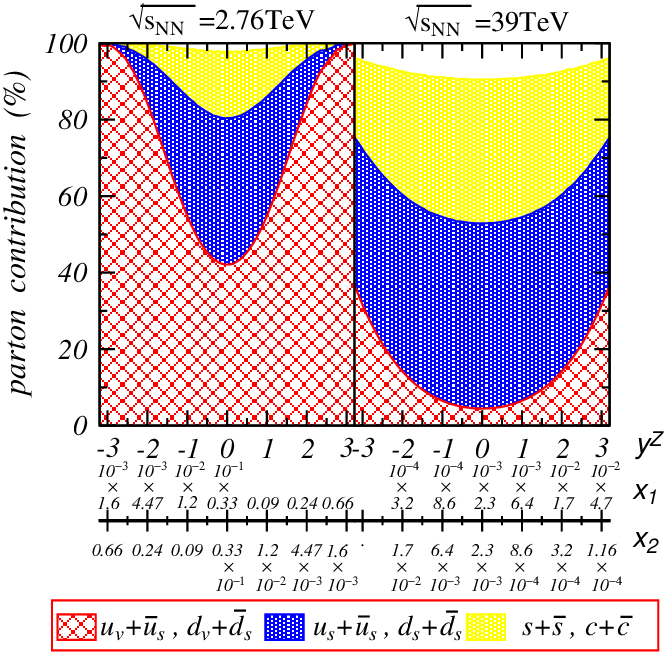}
\caption{(Color online) The contribution rates of various partonic subprocess as functions of $y^Z$ in Pb+Pb collisions at
both $\sqrt{s_{NN}}=2.76$~TeV and $39$~TeV calculated at the leading order.}
\label{fig:Z-Rprocessy@Pb+Pb}
\end{figure}

We have mentioned that at LO $R_{AA}$ as a function of the vector boson rapidity
mainly depends on the nuclear modification on the quark and anti-quark distributions~\cite{Ru:2014yma}, considering the LO process $q\bar{q}\rightarrow V$.
Actually anti-quarks only come from the nucleon sea, while quarks could be the valence or from the sea.
To better understand the contributions to $R_{AA}$ given by some certain flavors, we calculate at LO the
contribution ratio of partonic subprocess to the differential cross section.
Three primary kinds of processes for both LHC and future HIC are shown in
Fig.~\ref{fig:Z-Rprocessy@Pb+Pb}: (1)~$u_v+\bar{u}_s$ and $d_v+\bar{d}_s$;
(2)~$u_s+\bar{u}_s$ and $d_s+\bar{d}_s$ and (3)~$s+\bar{s}$ and $c+\bar{c}$.
Processes~(2-3) completely depend on the sea quarks, while the process~(1) contains valence quarks.
Obvious difference could be seen between the two energies.
At LHC energy, the processes with valence quarks in the initial-state dominate the $Z^0$ production.
However in the future HIC their contribution will be reduced significantly, while the processes with pure-sea quarks in the initial-state
become more important.
The results demonstrate that sea quarks play an important role in the $Z^0$ boson production, especially at the future HIC.
A main reason is that, the sea quark densities increase very fast with the decreasing momentum fraction $x$, whereas the valence quark densities decrease sharply (more discussion could be seen in the \ref{section:appendix}).
We show at the bottom of Fig.~\ref{fig:Z-Rprocessy@Pb+Pb} the corresponding momentum fractions of the incoming
partons~($x_1$: forward moving; $x_2$: backward moving).
The strong suppression of $Z^0$ production in mid-rapidity region at the future HIC is lead by the shadowed nucleus sea quarks.
It is noted that the nuclear modification of gluon density will also give contribution through high order corrections.

We note that at the small-$x$ region the uncertainties of nPDFs are rather large,
thus the $R_{AA}$ given by different nPDFs may even merge with each other if error bar is included.
That means the nuclear effect is very unclear at small-$x$, and the future Pb+Pb collisions will provide invaluable information to constrain the nPDFs for sea quarks and gluons in the small-$x$.
Especially, the nuclear variations of $s$ and $c$ quarks densities, which has been constrained loosely by vector boson production at the LHC, should become important at the future HIC.
\begin{figure*}[t]
\includegraphics[scale=0.7]{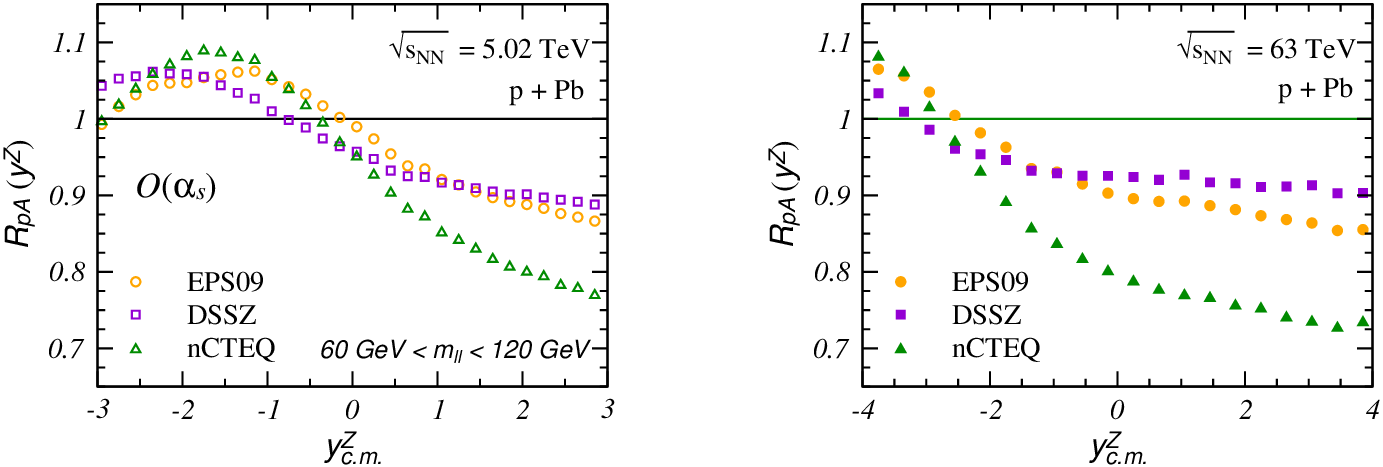}
\caption{(Color online) The nuclear modification factor $R_{pA}(y^Z)$ for p+Pb collisions at
$\sqrt{s_{NN}}=5.02$~TeV~(LHC, left panel) and $\sqrt{s_{NN}}=63$~TeV~(future HIC, right panel) at the order $\mathcal {O}(\alpha_s)$.}
\label{fig:Z-rpay@pPb}
\end{figure*}

\begin{figure}[b]
\includegraphics[scale=0.7]{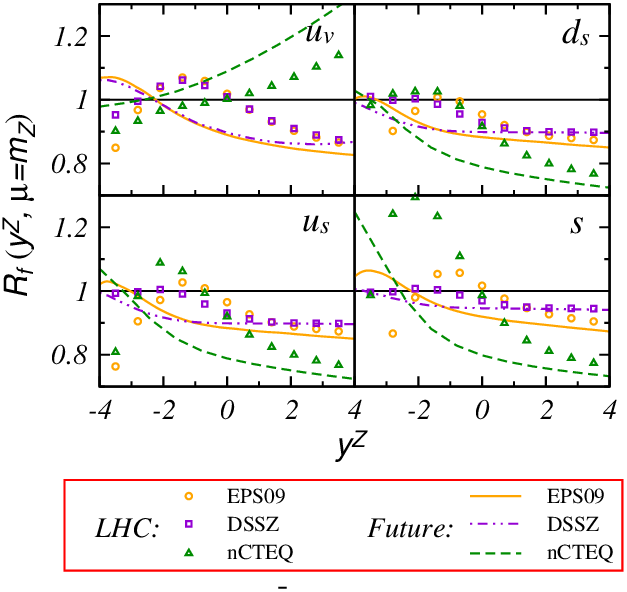}
\caption{(Color online) The flavor dependent nPDFs factors $R_f(y^Z,\mu)$ for p+Pb collisions
 at both $\sqrt{s_{NN}}=5.02$~TeV and $63$~TeV with factorization scale set at $m_Z$.}
\label{fig:Z-rfy@p+Pb}
\end{figure}

\begin{figure}[b]
\includegraphics[scale=0.7]{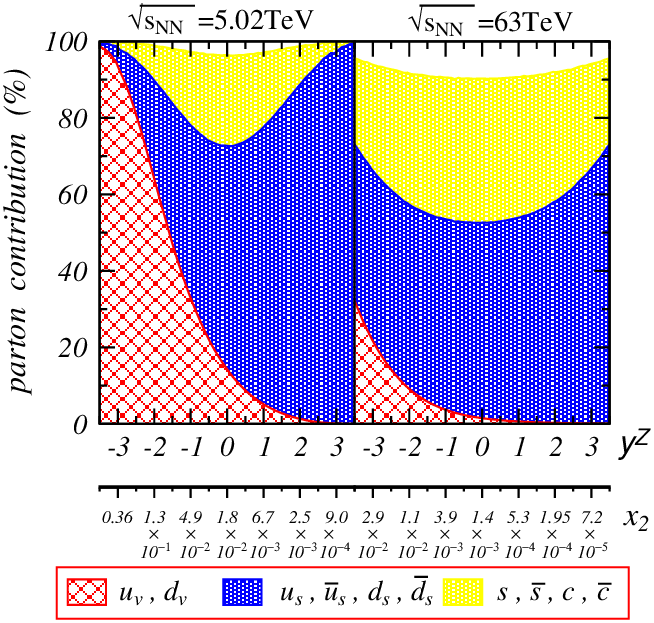}
\caption{(Color online) The LO rapidity dependent contribution rates for partonic subprocess initiated with various nuclear flavors in p+Pb collisions at
both $\sqrt{s_{NN}}=5.02$~TeV and $63$~TeV.}
\label{fig:Z-Rprocessy@p+Pb}
\end{figure}

The yield of $Z^0$ boson in the future p+Pb collisions at $\sqrt{s_{NN}}=63$~TeV is also studied.
Figure~\ref{fig:Z-rpay@pPb} shows the nuclear modification factor $R_{pA}$ as a function of the
$Z^0$ rapidity in the center of mass frame, predicted by EPS09, DSSZ and nCTEQ nPDFs.
Results for LHC energy~($\sqrt{s_{NN}}=5.02$~TeV) are also shown for comparison.
Since the nuclear effects come from the colliding lead nucleus, the asymmetric rapidity dependence of $R_{pA}$
could be observed at both LHC~(left panel) and future HIC~(right panel) energies.
For the LHC energy, enhancement can be seen in the backward region,
while suppression is given in the forward region~(related to anti-shadowing and shadowing, respectively~\cite{Ru:2014yma}).
For future energy, except the enhancements in very backward regime~($-4.0<y^Z<-3.0$), persistent suppressions are given by EPS09, DSSZ, as well as nCTEQ.

The LO analysis is simple for p+Pb collisions, and one could study the $R_{pA}$ from the nPDFs factors $R_f(y^Z,\mu)$ by
replacing the momentum fraction $x_2$~(related to the nuclear parton) with $y^Z$, according to Eq.~(\ref{xy}).
In Fig.~\ref{fig:Z-rfy@p+Pb} we plot the $R_f(y^Z,\mu)$ factors of certain flavors for both LHC and future HIC.
Compared to the LHC energy, the future p+Pb collisions generate more suppression in the rapidity region $y^Z\gtrsim-2.0$ due to
the shadowed initial parton from the lead nucleus with smaller-$x$.
 Though for $s$ quark density in the LHC backward region, and $u_v$ quark density in forward region of both energies,  the nuclear modifications with nCTEQ show distinct behaviour, their contributions to $Z^0$ rapidity distribution will be rather small in the corresponding region.
To see that clearly, three kinds of contribution of the partonic subprocess initiated by certain nuclear flavors are illustrated in Fig.~\ref{fig:Z-Rprocessy@p+Pb}.
It is observed that for both LHC and future HIC, the valence quarks'~($u_v$ and $d_v$) contribution decreases with the rapidity~(smaller-$x$),
but the sea quarks' increases.
Comparison between the two energies implies that the nucleus valence quarks densities give negligible contributions with the increasing energy~(smaller-$x$).
The suppression in the future p+Pb collisions is to a great extent the results of the shadowing effect on the sea quarks.
The contribution of the $s$ and $c$ quarks will also become non-trivial~(more than $30\%$ in the mid-rapidity at LO).
The $R_{AA}$ and $R_{pA}$ as functions of the $Z^0$ rapidity would become a good probe of the nuclear modification on small-$x$ sea
quarks in future heavy-ion collisions.
\begin{figure*}[t]
\includegraphics[scale=0.7]{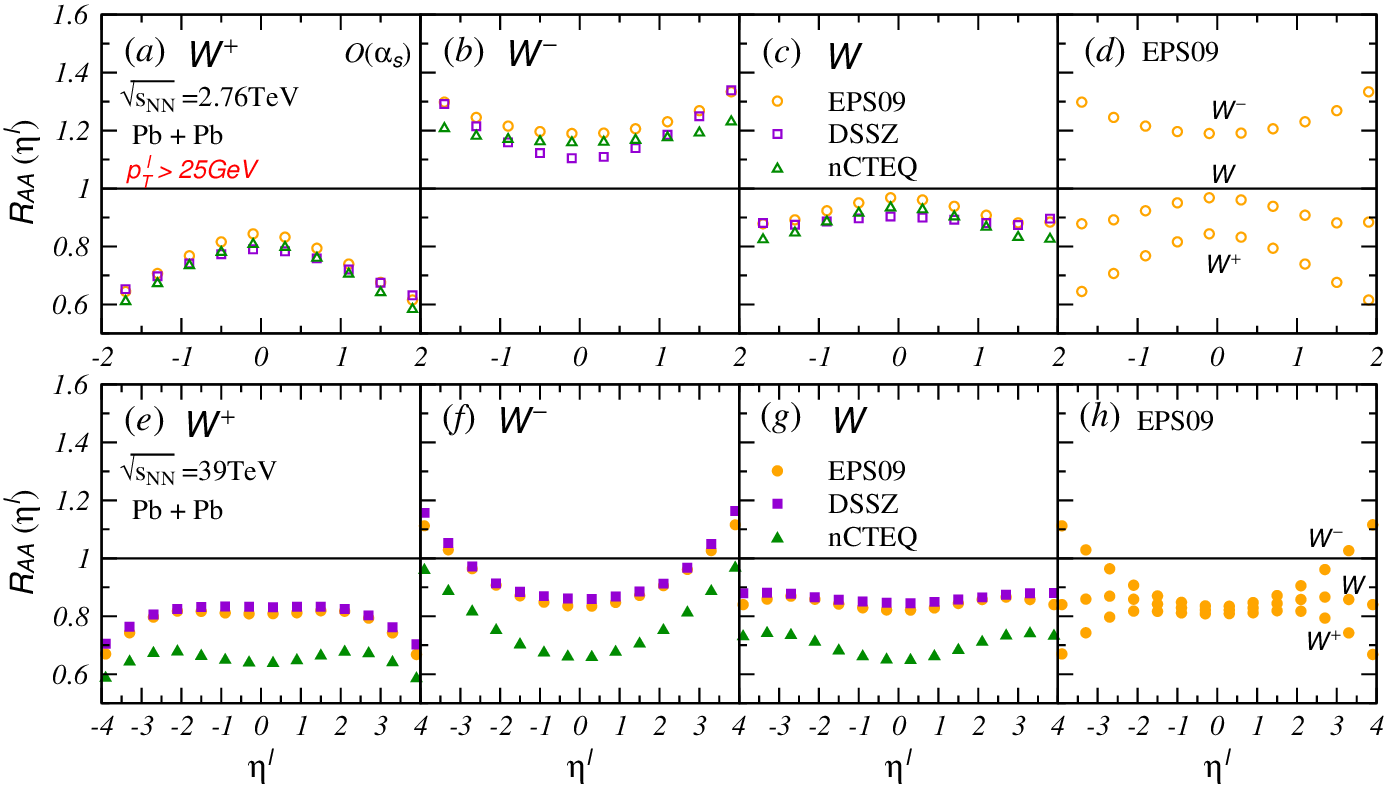}
\caption{(Color online) The nuclear modification factor $R_{AA}(\eta^l)$ for $W$ boson production in Pb+Pb collisions at
$\sqrt{s_{NN}}=2.76$~TeV~(LHC, top panel) and $\sqrt{s_{NN}}=39$~TeV~(future HIC, bottom panel) at the order $\mathcal {O}(\alpha_s)$.}
\label{fig:W-raaeta@PbPb}
\end{figure*}

\begin{figure*}[t]
\includegraphics[scale=0.7]{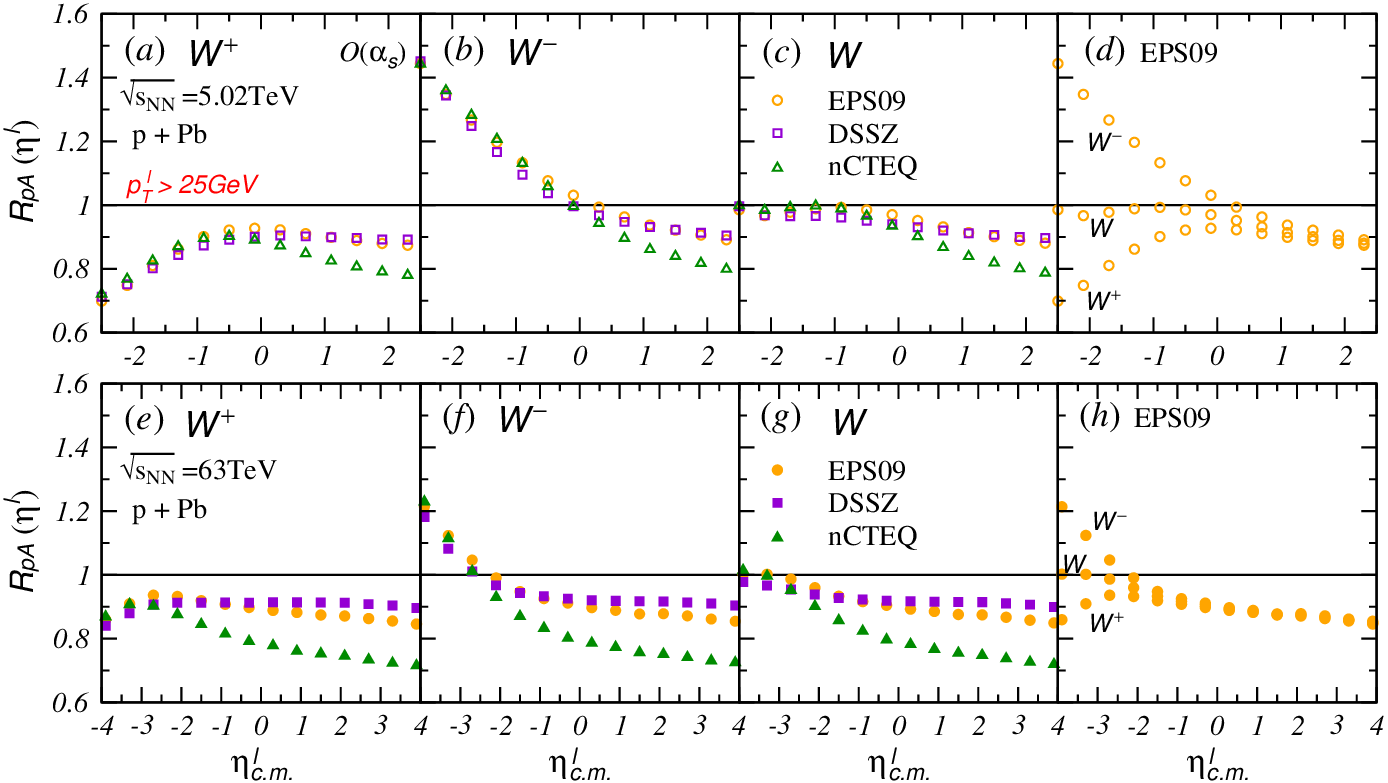}
\caption{(Color online) The nuclear modification factor $R_{pA}(\eta^l)$ for $W$ boson production in p+Pb collisions at
$\sqrt{s_{NN}}=5.02$~TeV~(LHC, top panel) and $\sqrt{s_{NN}}=63$~TeV~(future HIC, bottom panel) at the order $\mathcal {O}(\alpha_s)$.}
\label{fig:W-rpaETA@pPb}
\end{figure*}

For completeness, we also study the nuclear modification on the $W$ boson production.
Usually, the charged lepton pseudo-rapidity, instead of the boson rapidity,
is measured for $W$ boson production in experiment with the neutrino in final state~($W\rightarrow l\nu$).
We calculate the $R_{AA}$ and $R_{pA}$ as functions of the charged lepton pseudo-rapidity for $W$ production in
Pb+Pb collisions at $\sqrt{s_{NN}}=39$~TeV and p+Pb collisions at $\sqrt{s_{NN}}=63$~TeV, respectively. The transverse momentum region of the charged lepton is chosen to be $p_T^l>25$~GeV~\cite{Chatrchyan:2012nt,CMS:2014kla}.

For nucleus-nucleus collisions~(Fig.~\ref{fig:W-raaeta@PbPb}), asymmetry between $R_{AA}(W^+)$ and $R_{AA}(W^-)$,
due to the nuclear isospin effect, can be observed.
To be specific, because of the existence of neutrons in nuclei, the nuclear reaction may enhance the production of $W^-$ and reduce the yield of $W^+$, relative to those in the proton-proton collisions.
From the panels~(d) and (h), one can see that the separation between $W^+$ and $W^-$ is large at the LHC,
and becomes small at the future HIC, especially in the mid-rapidity region.
Actually, isospin effect is related to the parton distribution asymmetry~($u(x)\neq d(x)$, and $\bar{u}(x)\neq\bar{d}(x)$ ),
which originates mainly from the asymmetric $u/d$ valence quark distributions in a nucleon~(($uud$) for proton, and ($udd$) for neutron).
As is discussed before, at the future energy the initial parton tends to come from the smaller-$x$ regime, where the sea quarks dominate the valence quarks.
Thus, at smaller-$x$ the isospin effect becomes weaker~(more detail can be found in \ref{section:appendix}), and nuclear modifications on $W^+$ and $W^-$ become more symmetric.
It is also noted that, the nuclear effect of the total $W$~($=W^++W^-$) production,
is very similar as that of the $Z^0$ rapidity dependence at the same collision energy.
The differences among the $R_{AA}$ with three nPDFs sets are also consistent, for various particles~($W$, $W^+$, $W^-$, and $Z^0$) at the same collision energy.
Three nPDFs sets predict similar nuclear modifications at the LHC~(panels~(a-c)),
whereas nCTEQ gives much stronger suppressions in the future HIC~(panels~(e-g)).
Since isospin effect is very weak at the central rapidity, the suppression on the $W$ boson production at the future HIC
is mainly due to the shadowing effect on the nuclear sea quarks.

For proton-nucleus collisions~(Fig.~\ref{fig:W-rpaETA@pPb}), the nuclear variation on $W^\pm$ is again the result
of the isospin effect versus (anti-)shadowing effect.
At the LHC, one can observe that the isospin effect results in a separation between $W^+$ and $W^-$ in
the backward direction~(larger $x$ of the nucleus), and shadowing effect suppresses their production in the forward region~(smaller $x$ of the nucleus).
However, the isospin effect is rather weak at the future energy, and suppressions on both $W^+$ and $W^-$ in a wide range can be observed.
\begin{figure}[b]
\includegraphics[scale=0.8]{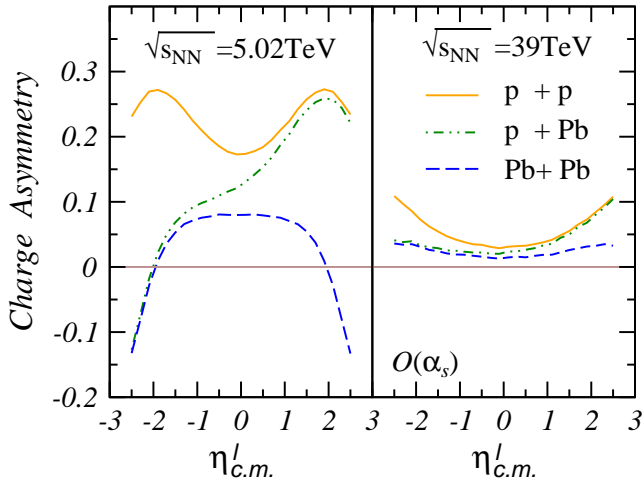}
\caption{(Color online) The charge asymmetry as a function of the charged lepton pseudo-rapidity in p+p, p+Pb and Pb+Pb
collisions for both the LHC~($\sqrt{s_{NN}}=5.02$~TeV) and future HIC~($\sqrt{s_{NN}}=39$~TeV).}
\label{fig:CASeta}
\end{figure}

The charge asymmetry observable $\mathcal {A}$ of $W^\pm$ production, defined as
\begin{eqnarray}
\label{casym}
\mathcal {A}=\frac{N_{W^+}-N_{W^-}}{N_{W^+}+N_{W^-}}.
\end{eqnarray}
is also studied for p+p, p+Pb and Pb+Pb collisions at both $\sqrt{s_{NN}}=5.02$~TeV and $\sqrt{s_{NN}}=3$9~TeV,
and are plotted in Fig.~\ref{fig:CASeta} as a function of the charged lepton pseudo-rapidity.
One can observe the charge asymmetries at the LHC show quite distinct behaviors in three different colliding systems, whereas the three curves at the future energy lie near the horizon and are close to each other.
It has been found that the nuclear variation of the charge asymmetry observable $\mathcal {A}$ of $W^\pm$ is mainly due to the isospin effect, and not sensitive to the other nuclear effects~\cite{Ru:2014yma}.
Therefore it is understandable that the isospin effect will become weak in the future HIC, as shown in Fig.~~\ref{fig:CASeta}.

\section{Nuclear modification ratio for vector boson transverse momentum distributions}
\label{section:ZWtransverse}

\begin{figure*}[t]
\includegraphics[scale=0.7]{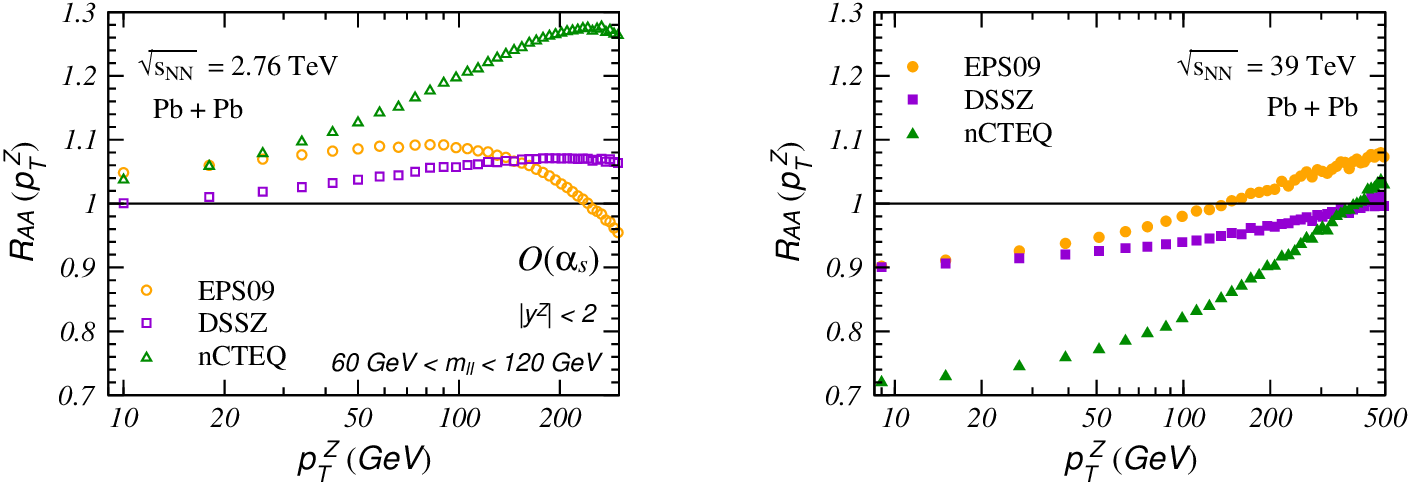}
\caption{(Color online) The nuclear modification ratio $R_{AA}$ of the $Z^0$ transverse momentum distribution in
Pb+Pb collisions at $\sqrt{s_{NN}}=2.76$~TeV~(LHC, left panel) and $\sqrt{s_{NN}}=39$~TeV~(future HIC, right panel) at the the order $\mathcal {O}(\alpha_s)$.}
\label{fig:Z-raapt@PbPb}
\end{figure*}

In this section we investigate the CNM effects on vector boson transverse momentum spectrum, which
may be used to get access to quite different kinematic regions from that by the boson rapidity production~\cite{Ru:2014yma}.
In this paper we mainly focus on the massive boson production with $p_T^V>10$~GeV, where DYNNLO provides an excellent baseline description
for $p_T^V$ distribution in elementary hadron-hadron collisions.

\begin{figure}[t]
\includegraphics[scale=0.7]{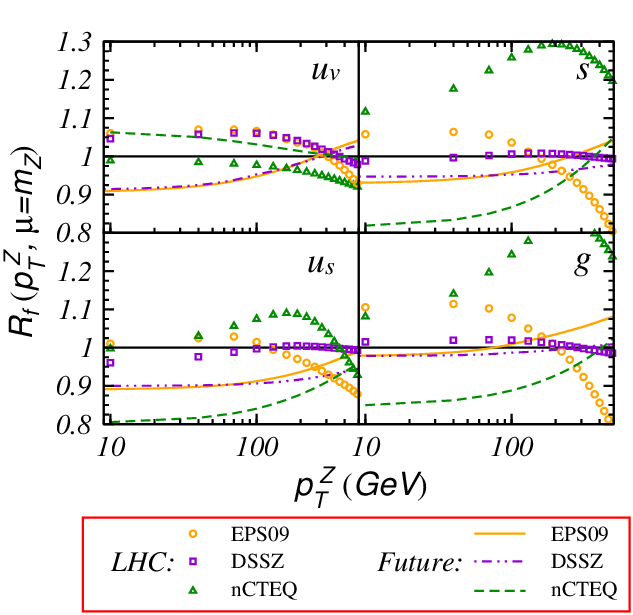}
\caption{(Color online) The flavor dependent nPDFs factors $R_f(p_T^Z,\mu)$ for Pb+Pb collisions at both $\sqrt{s_{NN}}=2.76$~TeV and $39$~TeV
with the factorization scale fixed at $m_Z$.}
\label{fig:Z-rfpt@Pb+Pb}
\end{figure}

\begin{figure}[t]
\includegraphics[scale=0.7]{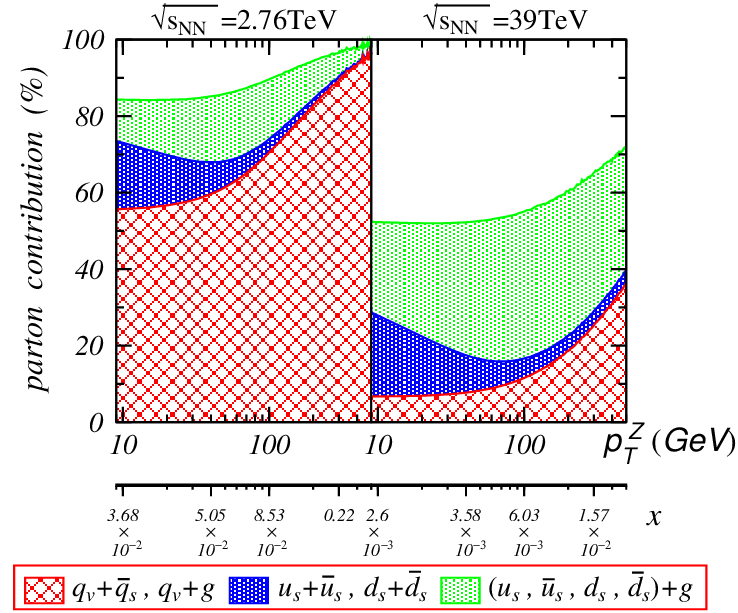}
\caption{(Color online) The $Z^0$ transverse momentum dependent contribution rates of various partonic subprocess in Pb+Pb collisions at both
$\sqrt{s_{NN}}=2.76$~TeV and $39$~TeV calculated at the LO.}
\label{fig:Z-Rprocesspt@Pb+Pb}
\end{figure}

First, we study the nuclear modification on $Z^0$ boson transverse momentum distribution in future Pb+Pb collisions at $\sqrt{s_{NN}}=39$~TeV.
Besides the same $Z^0$ invariant mass interval as used in the last section, the rapidity range $|y^Z|<2.0$ is also chosen.
The factor $R_{AA}$, as a function of $Z^0$ boson transverse momentum $p_T^Z$, is calculated at the order $\mathcal {O}(\alpha_s)$.
Although the differential cross section will be changed a little by higher order corrections,
$R_{AA}$ defined as the ratio of the cross section of A+A to that of p+p would not be very sensitive to the higher order corrections or the variation of hard scales~\cite{Vogt:2000hp,Ru:2014yma}.
Predictions by EPS09, DSSZ, and nCTEQ, for both LHC and future HIC, are shown in Fig.~\ref{fig:Z-raapt@PbPb}.
Obvious difference between the two collision energies can be seen.
Appreciable enhancements are given by three nPDFs sets in the region $10$~GeV$\lesssim p_T^Z\lesssim250$~GeV at the LHC,
whereas considerable suppressions are given by them in the region~$10$~GeV$\lesssim p_T^Z\lesssim150$~GeV for the future Pb+Pb collisions.
The nCTEQ nPDFs supports stronger nuclear modifications than the other two in the studied $p_T$ regime.

To well understand the differences among the nPDFs sets and also between the two collision energies, one can again perform the analysis at LO.
At this order, we obtain the kinematic relation
\begin{eqnarray}
\label{xpt}
x_{1,2}=\frac{p_T+\sqrt{p_T^2+m_{V}^2}}{\sqrt{s_{NN}}}
\end{eqnarray}
at mid-rapidity with the narrow width approximation.
Then according to Eq.~(\ref{xpt}), we replace the initial parton momentum fraction $x$ with $p_T^Z$ and plot the flavor dependent
factor $R_f(p_T^Z,\mu)$ of nPDFs in the Fig.~\ref{fig:Z-rfpt@Pb+Pb}.
For both LHC and future HIC, those factors of EPS09, DSSZ, and nCTEQ are shown.
Compared to the nuclear effect at the LHC, more suppression on the parton distribution could be seen in the region
$p_T^Z\lesssim200$~GeV for the future nucleus-nucleus collisions, except for the nCTEQ $u$ valence quark distribution.
Because the related initial parton momentum fractions are different~(smaller at future energy), shadowing effect dominates the future Pb+Pb collisions whereas anti-shadowing effect is shown up at the LHC.
\begin{figure*}[t]
\includegraphics[scale=0.7]{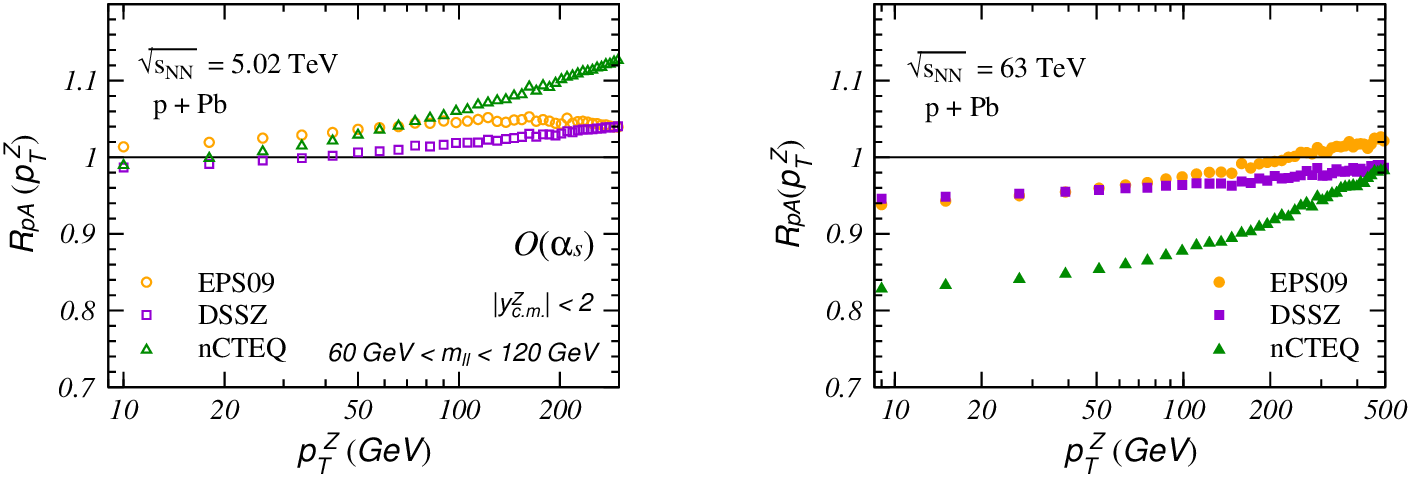}
\caption{(Color online) The nuclear modification factor $R_{pA}$ of the $Z^0$ boson transverse momentum spectrum for p+Pb
collisions at $\sqrt{s_{NN}}=5.02$~TeV~(LHC, left panel) and $\sqrt{s_{NN}}=63$~TeV~(future HIC, right panel) at the order $\mathcal {O}(\alpha_s)$.}
\label{fig:Z-rpapt@pPb}
\end{figure*}

To see clearly how the nuclear modification on each flavor comes into the observable $R_{AA}(p_T^Z)$, we again calculate the the contribution rates of the various partonic subprocesses for both the LHC and future HIC.
Three kinds of contribution are plotted in Fig.~\ref{fig:Z-Rprocesspt@Pb+Pb}:
(1)~Partonic subprocesses with valence quarks in the initial-state, such as $u_v+\bar{u}_s$ or $d_v+g$;
(2)~Pure $u,d$ sea initiated subprocesses as $u_s+\bar{u}_s$ and $d_s+\bar{d}_s$; and
(3)~Gluon initiated partonic processes with $u,d$ sea like $u_s+g$.
One could observe that processes of Type~(1) give the dominant contributions at the LHC energy, but their contributions decline significantly in the future HIC.
On the other hand, processes of Type (2) and Type (3) and other sea quark initiated processes dominate in the future energy collisions.
In the previous work~\cite{Kang:2012am,Ru:2014yma} on the nuclear effects on $p_T$ spectra of gauge bosons at LHC, it has been shown that the gluon's contribution is predominant.
Here we demonstrate that the impact of the nuclear modification of the sea quark distribution, on the $R_{AA}(p_T^Z)$, is comparable to that of the gluon, in the future nucleus-nucleus collisions.
For example, the large enhancement in the region $100$~GeV$\lesssim p_T^Z\lesssim300$~GeV at the LHC with nCTEQ,
is mainly due to the gluon nuclear modification but not the $s$ quark~(also enhanced a lot by nCTEQ), while at future HIC the shadowing of both gluon density and sea quark~($u,d,s,c$) distributions in nCTEQ give significant suppressions to $p_T$ distributions of $Z^0$.

\begin{figure}[t]
\includegraphics[scale=0.7]{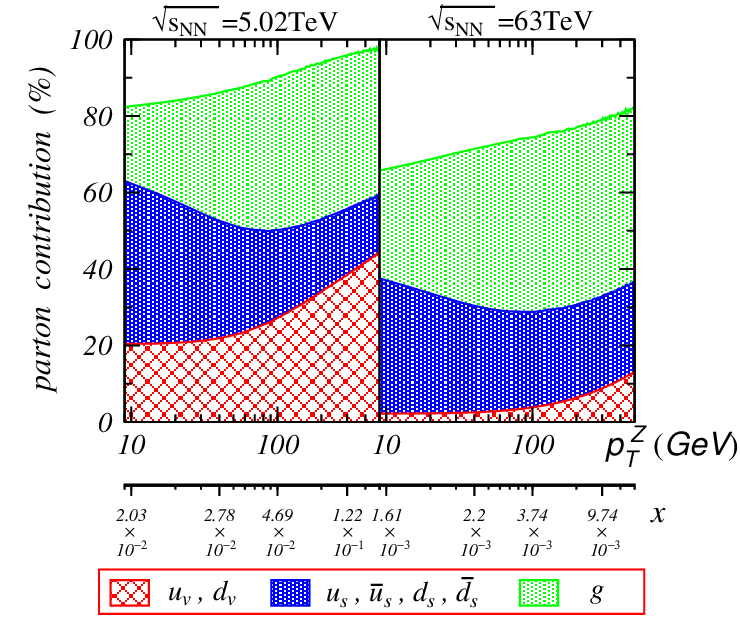}
\caption{(Color online) The LO partonic subprocess contribution rates as functions of $p_T^Z$ for different
nuclear flavors participating in p+Pb collisions at both $\sqrt{s_{NN}}=5.02$~TeV and $63$~TeV.}
\label{fig:Z-Rprocesspt@p+Pb}
\end{figure}

The nuclear modification factor of the $Z^0$ boson transverse momentum distribution in the future p+Pb collisions at $\sqrt{s_{NN}}=63$~TeV is also calculated at $\mathcal {O}(\alpha_s)$, as shown in Fig.~\ref{fig:Z-rpapt@pPb}.
Similar to that in Pb+Pb collisions, the suppression due to  the shadowing effect could be observed at the p+Pb in the future HIC.
The nuclear modification in p+Pb collisions is weaker than that in Pb+Pb collisions because in p+Pb only one colliding object is lead nucleus.
The partonic subprocess contribution rates are calculated for both LHC and future p+Pb reactions.
Fig.~\ref{fig:Z-Rprocesspt@p+Pb} shows the rates for three kinds of process, each of which is initiated by certain flavors in the lead nucleus.
One can see that for both of the two energies, processes initiated with sea quarks or gluons give the main contribution. Especially at future energy
the valence quarks' contribution is marginal, the nuclear effects are brought in by the sea quarks and gluon.

\begin{figure*}[t]
\includegraphics[scale=0.7]{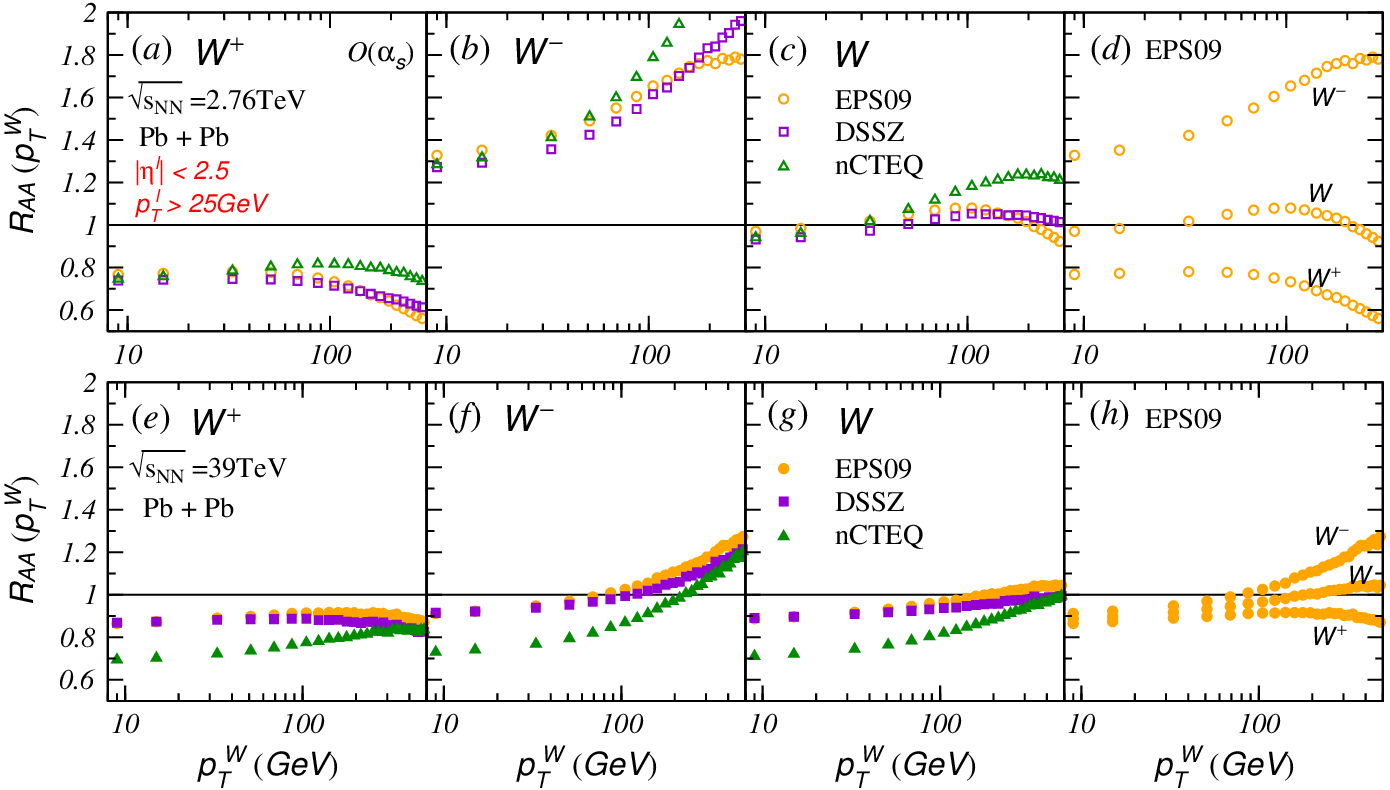}
\caption{(Color online) The nuclear modification ratio $R_{AA}(p_T^W)$ for $W^+$, $W^-$ and total $W(=W^++W^-)$ productions in the
Pb+Pb collisions at $\sqrt{s_{NN}}=2.76$~TeV~(LHC, top panel) and $\sqrt{s_{NN}}=39$~TeV~(future HIC, bottom panel) at the order $\mathcal {O}(\alpha_s)$.}
\label{fig:W-raapt@PbPb}
\end{figure*}
\begin{figure*}[t]
\includegraphics[scale=0.7]{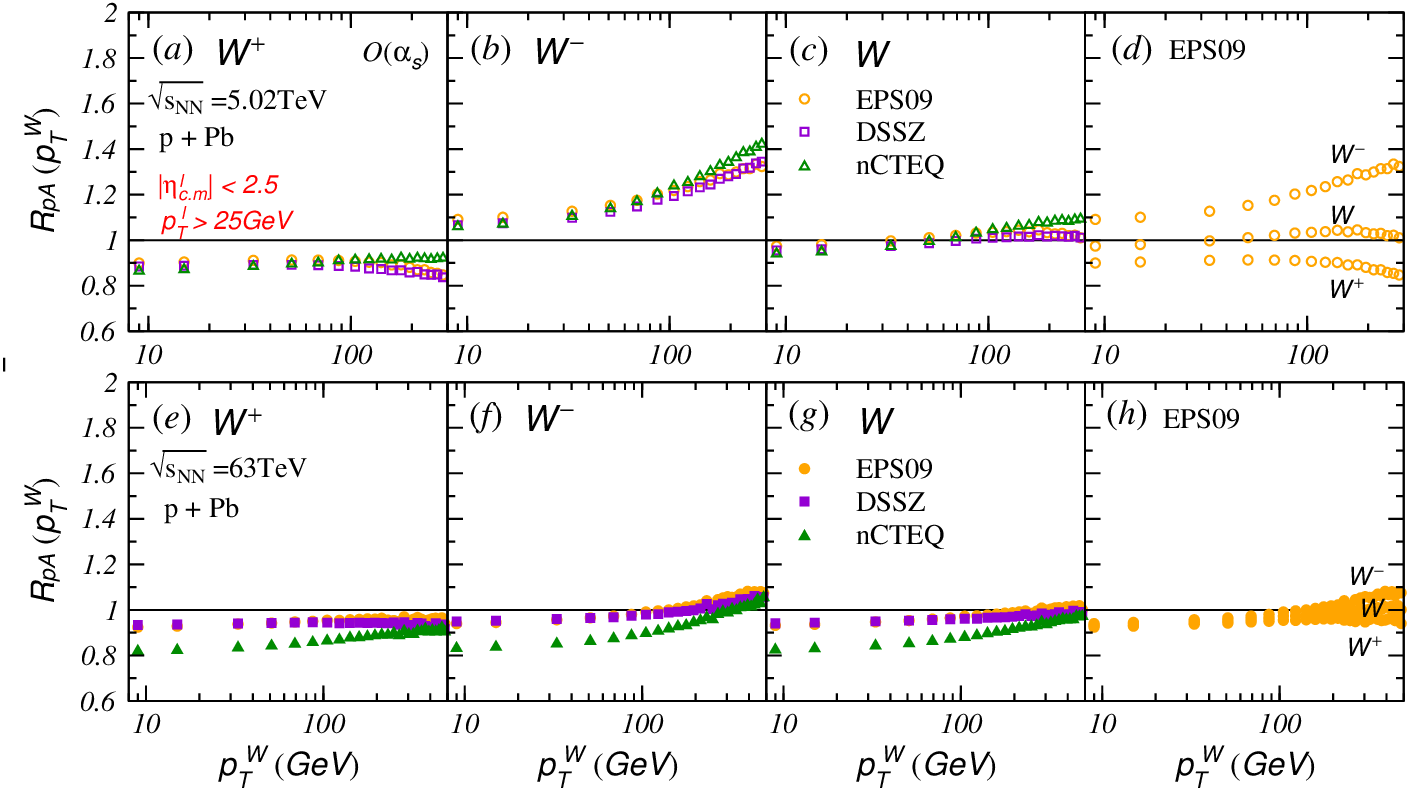}
\caption{(Color online) The nuclear modification factor $R_{pA}$ as a function of the $W$ boson transverse momentum for the p+Pb
collisions at $\sqrt{s_{NN}}=5.02$~TeV~(LHC, top panel) and$\sqrt{s_{NN}}=63$~TeV~(future HIC, bottom panel) at the order $\mathcal {O}(\alpha_s)$.}
\label{fig:W-rpapt@pPb}
\end{figure*}

Next we turn to study the nuclear modification on the $W$ boson transverse momentum spectrum.
The $W$ boson production is calculated in the final state phase space defined by the
center of mass frame pseudorapidity and transverse momentum of the charged lepton as
$|\eta^l|<2.5$ and $p_T^l<25$~GeV~\cite{CMS:2014kla}.
In Fig.~\ref{fig:W-raapt@PbPb} we plot $R_{AA}(p_T^W)$ in Pb+Pb collisions at both
$\sqrt{s_{NN}}=2.76$~TeV and $\sqrt{s_{NN}}=39$~TeV at the order $\mathcal {O}(\alpha_s)$.
The curves for $W^+$ and $W^-$ are separated from each other by the isospin effect, between which lies $R_{AA}$ for total $W$~($=W^++W^-$).
The separation increases with the $W$ transverse momentum~(increasing with $x$), and decreases with collision energy~(decreasing with $x$).
Although $W^\pm$ production obviously depends on the isospin effect, that of total $W$ receives
a $Z^0$-like nuclear variation mainly because of the (anti-)shadowing effect.
One can observe the $R_{AA}$ of $W^-$ boson are very different for two collision energies.
It is because that the isospin effect strongly enhances the $W^-$ production at the LHC, while the
shadowing effect is more pronounced than the isospin effect in the region $p_T^W<100$~GeV at the future HIC.
It should also be noted that, the suppression on $W^+$ at the LHC is due to the isospin effect,
but that at the future energy is caused by the shadowing effect.

The nuclear modification ratio $R_{pA}(p_T^W)$ is also calculated for the future p+Pb collisions at $\sqrt{s_{NN}}=63$~TeV,
and the results are shown in Fig.~\ref{fig:W-rpapt@pPb}.
Results for the LHC energy are also shown for comparison.
Same trends as in Pb+Pb collisions could be seen in p+Pb.
It should be noted that although both isospin effect and (anti-)shadowing effect are weaker in p+Pb as compared to Pb+Pb, the shadowing effect still surpasses the isospin effect
in the $p_T^W<100$~GeV region and thus suppresses the $W^-$ yield in the future HIC.

\begin{figure}[t]
\includegraphics[scale=0.7]{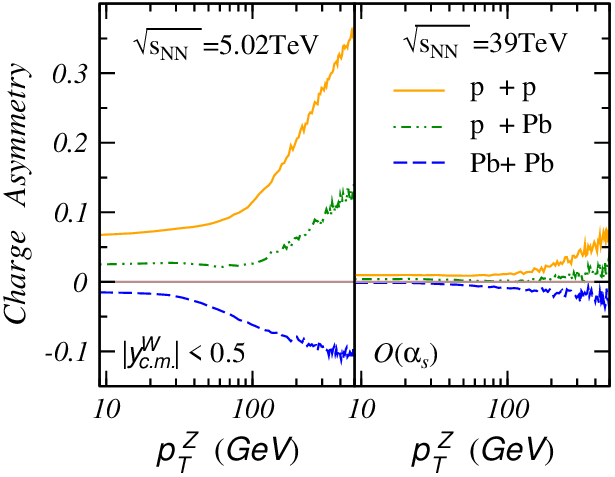}
\caption{(Color online) The charge asymmetry as a function of the $W$ boson transverse momentum in p+p, p+Pb and Pb+Pb
collisions for both the LHC~($\sqrt{s_{NN}}=5.02$~TeV) and future HIC~($\sqrt{s_{NN}}=39$~TeV).}
\label{fig:CASpt}
\end{figure}

Next we study the charge asymmetry observable $\mathcal {A}$ as a function of the $W$ transverse momentum.
The results for p+p, p+Pb and Pb+Pb collisions at both $\sqrt{s_{NN}}=5.02$~TeV and $\sqrt{s_{NN}}=3$9~TeV
are plotted in Fig.~\ref{fig:CASpt}.
One can observe that the nuclear modification of $\mathcal {A}$ due to isospin effect increases
with $p_T^W$ and decreases with the collision energy.
Actually $\mathcal {A}$'s nuclear modification also increases with its value in proton-proton collisions which is related to the
parton distribution asymmetry ratio $r_{ud}(x)$ and $r_{\bar{u}\bar{d}}(x)$~(definition can be found in \ref{section:appendix}).
Figure.~\ref{fig:CASpt} demonstrates the isospin effect will be reduced in the future HIC.

\begin{figure}[t]
\includegraphics[scale=0.7]{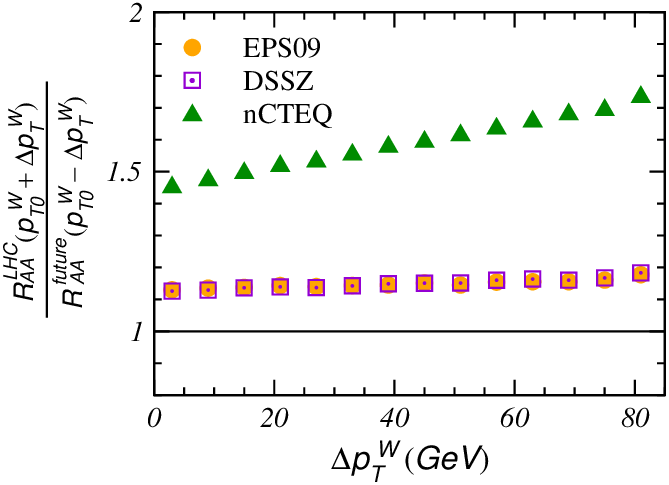}
\caption{(Color online) The ratio $\mathcal {R}(\Delta p_T^W)$ of the $R_{AA}(p_T^W)$ at $\sqrt{s_{NN}}=2.76$~TeV to
that at $\sqrt{s_{NN}}=39$~TeV for total $W$ production.}
\label{fig:W-raampt@pPb}
\end{figure}

From the numerical results we note that the differences among various nowadays nPDFs sets are considerable.
As an example, we calculate the ratio of the $R_{AA}(p_T^W)$ at the LHC~($\sqrt{s_{NN}}=2.76$~TeV) to that at future energy~($\sqrt{s_{NN}}=39$~TeV) for total
$W$ production, as
\begin{eqnarray}
\label{raam}
\mathcal {R}(\Delta p_T^W)=\frac{R_{AA}^{LHC}(p_{T0}^W+\Delta p_T^W)}{R_{AA}^{Futu}(p_{T0}^W-\Delta p_T^W)}.
\end{eqnarray}
The results for $p_{T0}^W=90$~GeV are shown in Fig.~\ref{fig:W-raampt@pPb}.
One can observe that EPS09 and DSSZ predict similar ratios of $R_{AA}$ at the LHC and future HIC, with very flat $\Delta p_T^W$ dependence.
However, nCTEQ gives a very different result: the ratio is obviously larger than those with the other two nPDFs, and increases faster with $\Delta p_T^W$. The measurement on nuclear modification of the vector boson production in the future heavy-ion collisions will provide very useful constraints on nPDFs.

\section{Summary}
\label{section:summary&conclusion}
We investigate in this paper the initial cold nuclear matter effects on the heavy vector boson production
in the future nuclear collisions.
Within the framework of perturbative QCD, we numerically calculate the nuclear modification ratios of the
$Z^0$ boson rapidity distribution, the charged lepton pseudo-rapidity dependence for $W$ production,
as well as $Z^0/W^\pm$ transverse momentum spectra.
Both Pb+Pb collisions at $\sqrt{s_{NN}}=39$~TeV and p+Pb collisions at $\sqrt{s_{NN}}=63$~TeV are studied,
where the CNM effects are included by utilizing the parameterized nuclear parton distribution functions sets~(EPS09, DSSZ and nCTEQ).

Different nuclear modifications from those at the LHC energies could be found.
First, the nuclear shadowing effect would play a more important role in massive gauge boson production
in the future HIC.
As consequences, both the (pseudo-)rapidity dependence and transverse momentum distribution of the boson would receive
suppressions in a wide regime.
Second, the nuclear isospin effect would be rather weak in the future nuclear reactions,
which could be clearly observed in the $W^\pm$ production.
For example, the $W^-$ production in the future HIC is even suppressed in the $p_T^W<100$~GeV region,
though at the LHC it is enhanced by isospin effect.

These very distinct nuclear modifications at the future HIC with those at the LHC result mainly from the smaller momentum
fraction $x$ carried by the initial partons at the fixed rapidity or $p_T$.
At the LHC, the momentum fraction of the initial-state partons may fall in the anti-shadowing region~(for EPS09), however in the future HIC it may be seen in the kinematic region of the shadowing effect.
Furthermore, the calculations of the contribution rates for various partonic subprocesses show that the valence quarks
gradually lose their relevance with the increasing colliding energy because of their sharply depleted densities in the smaller-$x$ region.
With the highly suppressed $u$ and $d$ valence quark distributions, the valence quark density asymmetry is minor, and the isospin effects
would become much weaker in the future HIC.

In addition, large differences among results with several parametrization sets of nPDFs could also be seen at the future HIC.
For example, results with nCTEQ predict more suppression for the vector boson yields than that with EPS09 and DSSZ in future HIC.
The theoretical understanding of nuclear modifications of PDFs in nuclear collisions is far from satisfactory, and
the massive gauge boson production in the future HIC could provide an excellent probe
of the shadowing effect on the nuclear sea quark distribution as well as gluon density by imposing new and precise constraints on the nPDFs
at rather small-$x$.

\begin{appendix}
\section{$C^f_{v,s}(x)$ and $r_{u,d}(x)/r_{\bar{u},\bar{d}}(x)$}
\label{section:appendix}
\begin{figure}[t]
\includegraphics[scale=0.7]{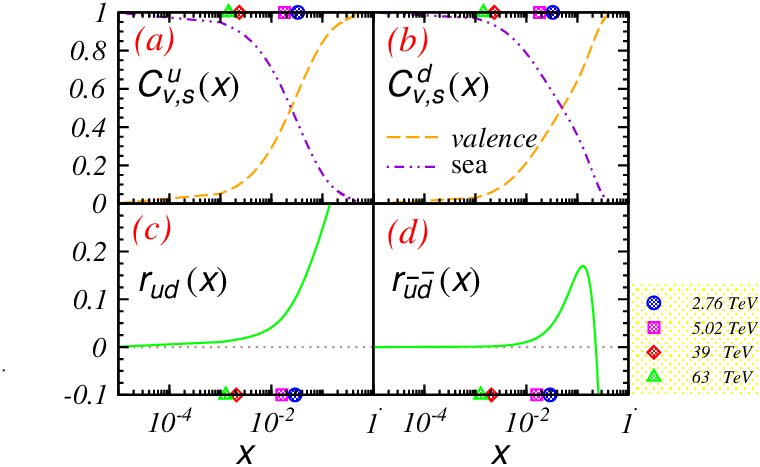}
\caption{(Color online) Panels~(a,b): the $C^f_{v,s}(x)$ of valence and sea for $u$ and $d$ quarks in a proton.
Panels~(c,d): the parton distribution asymmetry ratio $r_{u,d}(x)$ and $r_{\bar{u},\bar{d}}(x)$.
The factorization scale is set at the $Z^0$ boson mass $m_Z$.
The marks at $x$-axis are corresponding to the different colliding energies. }
\label{fig:Cvs&Rud}
\end{figure}

We calculate the $C^f_{v,s}(x)$ of the valence and sea quarks in a proton with the MSTW2008 PDFs~\cite{Martin:2009iq} as
\begin{eqnarray}
\label{CVS}
C^f_v(x)&=&\frac{f_v(x)}{f_v(x)+f_s(x)} \nonumber \\
C^f_s(x)&=&\frac{f_s(x)}{f_v(x)+f_s(x)}=1-C^f_v(x),
\end{eqnarray}
where $f=u,d$.
The results of $u$ and $d$ quarks are plotted in the panels~(a) and (b) of Fig.~\ref{fig:Cvs&Rud}, respectively.
One can see that for both $u$ and $d$, valence quark $C^f_{v}(x)$ increase with the parton momentum fraction $x$, and decrease
with the collision energy.

The parton distribution asymmetry $r_{u,d}(x)$ and $r_{\bar{u},\bar{d}}(x)$ defined as
\begin{eqnarray}
\label{Rudusds}
r_{ud}(x)&=&\frac{u(x)-d(x)}{u(x)+d(x)}, \nonumber \\
r_{\bar{u}\bar{d}}(x)&=&-\frac{\bar{u}(x)-\bar{d}(x)}{\bar{u}(x)+\bar{d}(x)}
\end{eqnarray}
are also plotted in the panels~(c) and (d) of Fig.~\ref{fig:Cvs&Rud}.
One could observe that the parton distribution asymmetries are small in the low-$x$ region.
As a result, the isospin effect would be moderate in that region.
We also mark at the $x$-axis the points corresponding to different energies according to the relation $x\sim m_V/\sqrt{s_{NN}}$.
The marked values in panels~(a) and (b) are obtain with $m_V=m_Z$ and those in (c) and (d)  with $m_V=m_W$.
The correlation between the $C^f_v(x)$ of valence quark and the parton asymmetry ratio
$r_{u,d}(x)$~(and $r_{\bar{u},\bar{d}}(x)$) should be emphasized.
Actually, the parton distribution asymmetry is to a large extent caused by the valence quark~($u_v$, $d_v$) density asymmetry, as a
result the tenuous valence quark density would give a much weaker $r_{ud}$ (or $r_{\bar{u}\bar{d}}$) in the small-$x$ regime.
\end{appendix}

\begin{acknowledgements}
P.~R would like to thank M.~Grazzini, Ning-Bo Chang, and Yuncun He for helpful discussions, and also thank Jing Yang for the help in computing techniques.
This research is supported in part by the MOST in China under Contract No. 2014CB845404,
and by Natural Science Foundation of China with Project Nos. 11322546,  11435004, 11221504, 11275037.
\end{acknowledgements}

\end{document}